# Education Projects for Sustainable Development: Evidence from Ural Federal University[1]

**Marina Volkova[1], Jol Stoffers[2] and Dmitry Kochetkov[3,4*]**

[1]Ural Federal University, Russian Federation; m.v.volkova@urfu.ru

[2]Research Centre for Employability, Zuyd University of Applied Sciences, The Netherlands; jol.stoffers@zuyd.nl

[3]RUDN University, Russian Federation; kochetkov-dm@rudn.ru

[4]Higher School of Economics, Russian Federation; d.kochetkov@hse.ru

*Correspondence: kochetkovdm@hotmail.com

**Abstract:** Sustainable development is a worldwide recognized social and political goal, discussed in both academic and political discourse and with much research on the topic related to sustainable development in higher education. Since mental models are formed more effectively at school age, we propose a new way of thinking that will help achieve this goal. This paper was written in the context of Russia, where the topic of sustainable development in education is poorly developed. The authors used the classical methodology of the case analysis. The analysis and interpretation of the results were conducted in the framework of the institutional theory. Presented is the case of Ural Federal University, which has been working for several years on the creation of a device for the purification of industrial sewer water in the framework of an initiative student group. Schoolchildren recently joined the program, and such projects have been called university-to-school projects. Successful solutions of inventive tasks contribute to the formation of mental models. This case has been analyzed in terms of institutionalism, and the authors argue for the primacy of mental institutions over normative ones during sustainable society construction. This case study is the first to analyze a partnership between a Federal University and local schools regarding sustainable education and proposes a new way of thinking.

1. Introduction

Two polar viewpoints exist regarding Earth's climate, with proponents and opponents of the concept of global warming agreeing that humanity has contributed to climate change. Researchers are paying more and more attention to studying the causes and effects of climate change from the standpoint of social sciences (Rosa and Dietz, 2012; Rosa *et al.*, 2015; Jorgenson *et al.*, 2019). The growth of the world's population and technological advancements have required increasing amounts of energy, which can be obtained only by burning fuel. Consequently, technogenic pollution of the planet has grown steadily over the years and air quality in large cities does not meet regulatory standards, leading to increases to respiratory, cardiovascular, and allergenic diseases and generating annual financial losses comparable to the GDP of a small country. Another prominent issue is water quality. The amount of water in nature is fixed, and for use in industry, agriculture, and everyday life, only fresh water is suitable. Contaminated water causes a decrease in the quality of drinking water, thermal pollution of water bodies, and flooding of territories. The demand for water has reached a level such that in many places, including Europe, there is an acute problem regarding the lack of fresh water. Scientists warn that in one or two generations, most of the world's population will lack fresh water (Connor, 2013).

---
[1] This study was first presented at the International Conference on Sustainable Cities (Sandler, Volkova and Kochetkov, 2018).

In a synthesis report, the United Nations (UN) Secretary-General stressed sustainable development goals (SDG) for everyone (United Nations, 2014) and that no one should be left behind (Chin and Jacobsson, 2016). Universities have demonstrated a trend of redesigning their strategies and organizations in line with principles of sustainability (Beynaghi *et al.*, 2016). Sustainability is then transformed from a component of education to a social learning process (Barth and Michelsen, 2013). Universities commonly join global networks such as the Sustainable Development Solutions Network, supported by the UN, the International Sustainable Campus Network (ISCN), the Association for the Advancement of Sustainability in Higher Education (AASHE) in the United States, and the Environmental Association for Universities and Colleges (EAUC) in the United Kingdom (Soini *et al.*, 2018). A modern interpretation of sustainable development includes environmental, economic, and social dimensions (Shriberg, 2004; Kuhlman and Farrington, 2010; Farley and Smith, 2013).

The UN's vision emphasizes the role of education in building the future (United Nations, 2015). Concerning education for sustainable development (ESD), this does not entail gathering recyclable materials, feeding birds in winter, and participation in ecological events. These are essential components, but not the most vital to the present argument. Several studies suggest that the growth of professional knowledge of a scientific, technological, and socioeconomic nature directly enhances an innovative development path (Barth and Michelsen, 2013). Thus, for the greening of production to occur, there must be an increase in the general culture of people in an area. The pivotal outcome of intellectualization of the population is the emergence of intellectual and innovative space. A new type of behavior is required for its functioning, not only for individuals but the social and economic systems as a whole.

Intellectualization of the population and various innovations arising from it cannot develop in a short time; any society needs a long period for its maturation, which includes the accumulation of knowledge and then turning it into a way of thinking that is suitable to the bulk of the population. Intellectual loners can only offer original ideas, describe them, and make prototypes, but only the collective efforts of a skilled workforce can transfer ideas into mass production. As a source of knowledge accumulation and generation, secondary and higher education moves to the forefront of the development of a sustainable society. Relevant skills and competencies are essential to implementing the paradigm of sustainable development (Bochko, 2015). Competence-oriented programs have been recognized as a factor in the transition of education in the direction of sustainable development (Jacobi, Toledo and Grandisoli, 2016), a concept that many empirical studies support (UE4SD, 2015; Dlouhá, Glavič and Barton, 2017).

Nevertheless, most discussions on the formation of competencies relevant to sustainable development are conducted in the context of higher education (Wiek, Withycombe and Redman, 2011; Lambrechts *et al.*, 2013; Mulà *et al.*, 2017). One viewpoint suggests that the agenda of sustainable development within a framework of educational discourse is controversial and problematic, with some authors arguing that education on sustainable development is a product and carrier of globalization (Jickling and Wals, 2008). Many universities throughout the world have adapted Sustainable Development as part of their mission; however, most SD efforts were not holistically integrated throughout the HEI system. The level of individual competences is well reviewed in (Lozano *et al.*, 2015).

Kasimov (2013) summarizes ESD in Russia as:

*The ideology of ESD in Russia as a whole is not denied, but it is not perceived as a universal educational paradigm that objectively reflects the challenges of the time. Promotion of ESD in Russia objectively corresponds to the interests of the country, but in fact, it is inhibited.*

The same is demonstrated by Ermakov (2013), who surveyed ESD experts, including researchers, university professors, methodologists, and activists of public organizations. The average assessment of the implementation of the UN's Economic Commission for Europe (ECE) strategy objectives for ESD in the Russian Federation (on a 4-point scale: 0=activity not started, 1=activity is conducted, 2=activity is developing, and 3=activity is completed) was only 0.5. A survey of teachers in several regions of Russia conducted by the same author demonstrates that 77.4% of respondents are aware of the need for ESD implementation in all educational institutions and at all levels of education. However, only a small portion (10.8%) perceive that their methodological training is sufficient for ESD implementation, and 30.4% found it difficult to answer, which most likely indicates poor familiarity with ESD (Ermakov, 2011).

Within the framework of technical education, teachers often treat sustainable development as an "aspect" of engineering (Mulder, Desha and Hargroves, 2013). Sustainability issues need to be addressed at the meta-level, using a holistic system approach, so that decisions can be made about conflicting issues arising from (sub)disciplinary analysis. This approach was implemented, for example, at the Energy Engineering Graduate Program at Başkent University (Wood and Edis, 2011). We argue that sustainable development requires not only skills and competencies but the formation of stable patterns of thinking. This is where secondary education comes to the forefront since the formation of mental models at an early age is much more effective. Children are required to study at school from 7 to 16 years. In junior classes, natural science lessons are introduced, where students are taught generally about the environment and ecology. There are also Olympiads of ecological directions for students. In some schools, children, under the guidance of teachers and supervisors, write reports or conduct feasibility studies, often an analysis of data from the Internet, on environmental topics. At university, this type of education ceases due to the organization of studies such that at the department level (apart from specialized ones), sustainable development and ecology are not primary subjects. Consequently, students have acquired some knowledge on the topic, but a way of thinking has not been formed. There is also currently lack of innovative breakthroughs on the topic of sustainable education.

Anthropogenic impacts on the Earth's biosphere is so global that governments of the world must cooperate in the field of ecology, though nuances are at play. A desire to reduce emissions of carbon dioxide into the atmosphere led to heated debates concerning the Paris Climate Agreement. One country might have concerns about reducing the anthropogenic impact on the planet, and another, for example, Great Britain, uses a climate agenda to protect its market. Some argue that global markets have a problem with carbon protectionism, which can lead to sanctions and pressure on national economies (Kilkiş, 2015). In this context, the withdrawal of the United States from the Paris Agreement represents merely protection of national business. Taking the problem more broadly, the entire history of the struggle for sustainable development is parallel to the history of the struggle of the population for the environment, whereas, for entrepreneurs, it is for profit. Pollution control facilities are often too expensive for medium and small businesses, leaving an uneven advantage for large plants and enterprises. Thus, the choice between profit and environment among most entrepreneurs in the world is obvious. The only deterrent to ecological irresponsibility is the regulatory role of the government regarding environmental protection with the rebuke of penalties, sanctions, etc. There are entrepreneurs for whom taking care of the environment is a priority, but for most, profit comes first and delegating environmental expenditures is an afterthought. The most environmentally conscious are young people who grew up and received education at a time when environmental problems became relevant and were discussed openly. As practice demonstrates, if a person has been exposed to environmental education since childhood, according to them, a businessperson and an ecologist should be able to make an inner compromise. As in most emerging countries, there is nearly no formal education policy concerning sustainable development in Russia. In this context, such projects are particularly important. We analyze the experiences of a partnership between Ural Federal University and local schools regarding sustainable education, and we call such projects university-to-school projects.

## 2. Recent Research

A detailed review of the literature is beyond the scope of this study, especially now a sufficient number of reviews in the field have been published. We can refer an interested reader, for example, to the work of Agbedahin (2019), which draws a certain line under empirical research and suggests future directions of research in the field of ESD. The work shows the relationship between Education for Sustainable Development (ESD) and the achievement of Sustainable Development Goals (SDG). Considerable attention was paid to the nexus between environmental education and ESD. Franco *et al.* (2018) analyze the practice of applying ESD in educational policies and curricula. The authors argued that the achievement of SDG largely depends on a better understanding of the existing gaps, focus areas, and regional differences between the Higher Education for Sustainable Development (HEfESD) policies. Sinakou *et al.* (2018) found that teachers do not take the concept of sustainable development holistically; usually, they put more emphasis on the economic and social aspects of SD. The authors of the article argued that teacher training programs should implement a holistic approach. The Russian researchers came to similar conclusions highlighting the following problems in the development of ESD (Grishaeva *et al.*, 2018):

- the discrepancy between the trends of higher education and the objective demands in society and politics that contradict sustainable development in general
- insufficient level of development of critical thinking of students in the process of environmental education as well as the pedagogical problem of the formation of students' reflective experience
- the problem of "locality" and "narrow targeting" of learning content
- the problem of the development of interdisciplinary and transdisciplinary approaches to environmental education
- the issue of pedagogical adaptation of axiological bases of ESD.

The study by Svanström *et al.* (2018) considered the possibility of using Multivariate Data Analysis (MVDA) and concept maps in the learning process. The work showed the relationship between personal characteristics and background of students on the one hand and performance in ESD on the other. The work by Molderez and Ceulemans (2018) highlighted systems thinking as a critical competence for ESD. Interestingly, the development of systems thinking, in this case, is built through the perception of the pieces of art. Molderez and Fonseca (2018) stressed the importance of real-world experience and service learning for the formation of sustainable development competencies.

## 3. Materials and Methods

We use the case study methodology as a research basis, a method that focuses on conducting intensive analyses of a situation that involves considering the context and using a combination of disparate research methods (i.e., qualitative and quantitative), data collection, and analysis. The method is described by Eisenhardt (1989) and Yin (1989), who propose an interpretation of the essence of a case study as a research strategy. Most contemporary researchers refer to Stake (Stake, 2000, 2008) and Yin (Yin, 2003, 2004), which became fundamental to subsequent studies based on the situational approach (Naumes and Naumes, 2006; Ellet, 2007; Easton, 2010). Researchers identify inherent problems with the case study method. Assessing higher education sustainability case studies, Corcoran, Walker and Wals (2004) discuss a gap between the internal need for contextual significance/importance and external demand for transferability/abstraction. Case studies are introspective and justified within a single institutional reality. If the purpose of a case study is to improve one's institutional practices, this might be sufficient, but if the purpose is also to improve institutional practices elsewhere, this should be reflected in the analysis structure. Therefore, during the current analysis, importance was placed on context, including the institutional landscape. Analysis of institutions enables a better understanding of the processes within a social system as a whole.

In recent decades, institutional analysis, sometimes differentiated during neo-institutional analyses, has developed strongly in numerous approaches. The tradition of the new institutional economics is based on writings from North (1990), Eccles and Williamson (1987), and others. In sociology, the contemporary institutional analysis is summarized by W. Scott, M. Brinton, V. Nee, P. Di Maggio, and W. Powell (Romanelli, Powell and DiMaggio, 1992; Scott, Brinton and Nee, 1999). Culture and symbolism are given much greater emphasis in institutional analysis than in standard, especially economic, analyses of organizations and behavior (Ostrom, 2010). We understand institutions as a set of formal norms, informal constraints, and coercive mechanisms (North, 1990). Institutions can be divided broadly into four groups:

- Normative – norms, rules, customs, standards, conventions, contracts, etc. (North, 1990);
- Functional – status functions and routines (Nelson, 1994; Searle, 1995);
- Structural – organizational forms and models of transactions (Scott, 1995);
- Mental – collective representations, beliefs, stereotypes, values, cognitive schemes, etc. (Denzau and North, 1994).

The study was grounded in the case of the Ural Federal University (Ekaterinburg, Russia); the case is described in the next section. The institutional analysis enables evaluation of the theoretical relevance of the case and its policy implications for Russia and the world.

**4. Results**

We argue that it is necessary to use school environmental practices during initial days of a project at a university, and essential to involve students when solving practical problems. Those who engaged in environmental problems at school take the work willingly. The task should not be too narrow, but rather large, incorporating knowledge from multiple areas. As an example, we cite the disposal and treatment of polluted waters of small combined heat and power stations (CHP) and thermal power plants (TPP). Main sources of hot water and heating are also primary origins of pollution in a living area due to their locations. Losses of heat supply from CHPs and TPPs increase depending on the distance to consumers. Historically, many of them are located in urban environments where it is impossible to place cooling ponds, sludge accumulators, etc. At the same time, a variety of heat exchangers, scrubbers, and other equipment can be used.

Due to technological processes, a large amount of contaminated water is emitted during steam condensation at a CHP or TPP. An even larger amount (10 to 15 times more than has been discharged) is needed to ensure that condensation and cooling of discharged water occur at the end. The most conservative estimates show that even a small CHP plant incurs significant losses as a result, and if the cost of water treatment and its disposal are added, the amount increases. CHPs and TPPs use water for washing boilers and other mechanisms, and if this water is thus returned to circulation for reuse, several problems can be solved simultaneously—reducing water consumption, reducing heat pollution, and obtaining an additional product—which will make the process more attractive to potential consumers and investors.

Such research has been conducted at Ural Federal University since 2008 under the supervision of M. Volkova, exclusively by students or volunteers who are not engaged in specialized departments. The problem involves a large range of issues, including the creation of a purification technique with the help of developed biofilters and the study of the complete use of filters. The team of students changed over the years, and the range of tasks varied. Each student can choose the direction in which they work. For example, construction of a prototype is more suited to builders, use and production to physicists, and biochemists might conduct a study of optimal conditions for the operation of biofilters. These students publish their findings, resulting in the formation of ecological thinking.

The team that is the focus of this study is uncommon; its composition changes regularly, all students are technical specialists, and the principle is voluntary work. Work in the research team occurs during the first year, and there is no compensation or funding. The team is working on a topical issue; the leader wants to show that environmental protection is not costly, as is commonly believed. Teams of similar-minded people wishing to improve the ecological situation are capable of solving environmental problems. The need for such work appears to be shaped by the educational component, in addition to the environmental one, and there was thus a desire to attract students to scientific work to broaden their horizons and develop their creative abilities. Involvement of students in solving practical environmental problems was effective. Students' engagement during their first year empowers a system of continuous environmental education. Creativity enables personal freedom, unites similar-minded people, and distracts from reality. Such teams are an excellent opportunity for the integration or inclusion of students with disabilities, who can realize their potential at creative discussions and while studying or processing experimental materials. If their state of health does not allow them to participate personally, modern means of communication allows these children to communicate remotely. For people with disabilities, much is done to create an accessible environment, but the most important aspect is a sense of necessity and usefulness to others, achieved by working in creative teams. It enables communication with others on equal footing, a realization of one's thoughts and projects, and in the end, raises self-esteem. We had an opportunity to verify this statement from experience. In a team of students during the Ural Project Session on Tavatue, there was a child with disabilities, expressed in the impossibility of concentrating continuously on work, quick fatigue, and absent-mindedness. He nevertheless gladly took readings from the instruments and researched information (for a longer period than the rest of the children). His role in preparing video reports for the presentation was irreplaceable due to his double role as both actor and director. Restrictions related to his health limitations did not allow him to engage in physical labor, but in the team, he was responsible for his part of the work.

What attracts students to a job that is not paid and is done in their spare time? According to the results of unstructured interviews with program participants [1], the answers are:

- a possibility to develop creative ideas;
- an opportunity to participate in the scientific conferences;
- the student might not become a scientist in the future, but an experience of presenting publicly is always useful.

The first presenters are usually hesitant, and the presentations are not gripping to the audience. However, after a year or two, the students are confident speakers, communicating effectively with the audience. The only thing that remains for students to remember about their work is certificates, presentations, and publications. From the data collected, the team of researchers developed a schematic diagram of the device, which has advantages such as compactness and wastelessness and can be adapted easily to particular features of production. Thus, from the original idea and due to the initiative of students and proposals that explore several aspects, the project divided into three parallel studies:

- application of purified water (for irrigation and plant cultivation);
- studying the possibility of obtaining biogas from biomass grown during the purification of condensate;
- development and improvement of the machine itself (patent documentation is being prepared).

A few issues require closer examination and study in the future. The team participated several times in competitions to receive funding from UMNIK [2], from which we obtained good reviews. We are hopeful that funding will come, and without it, all projects will have to be postponed and possibly ended. Fortunately, there are many creative children in the country, and they have a desire to contribute to environmental improvement. While working on the project and participating in conferences, the students unconsciously attracted attention to environmental issues. For this work and during different years, two received Vernadsky nominal scholarships [3]. Even more important are the social skills the students gained such that whatever they do in their lives, they will be able to rally themselves around like-minded people. Interesting studies and substantial proposals were developed, with enthusiasm as the sole driver. The students were given an opportunity to realize their ideas while doing something interesting and necessary. Not all were enrolled in post-graduate or master's studies in an environmental field. From custom, we preserve the work of previous generations of student volunteers in memory of the crucial work they performed. The quantity is irrelevant; the purpose is that students have a clearer understanding of the environment as a science. Some students continued their education, and the articles helped with admission to post-graduate or master's programs, and others found employment. Only 20% were enrolled in environmental master's studies, with one individual in postgraduate studies. Sixty percent majored in life safety [4] and chose an environmental career path after graduation.

Students can begin this work even earlier. Beginning in grades 7/8, a sufficient stock of knowledge accumulates among students, and their desires to apply them to practical tasks. A striking example is the Ural Project Session held for gifted students by Ural Federal University. To participate in the session, a student must pass a contest to which students from the 7th to 11th grades were admitted. The work was conducted in several directions, one of which was Intellectual Energy Systems. During the session, a project titled *Purification of Technical Water from Power Plants Using Simple Algae* was developed. The team consisted of 5 people—2 students from the 11th grade, two from 10th, and one from 8th. The students not only understood the problem perfectly but developed and assembled a prototype. The device consisted of three filters (Figure). Contaminated water entering the pipe in the first filter (far left) is subjected to primary purification from coarse contamination. In filters 2 and 3, there are algae, in which polluted water is saturated with oxygen, which changes the hardness. Indications of temperature, light, and hardness are controlled automatically by sensors on a control panel. If a sensor detects a decrease in hardness to a predetermined degree, the valve opens automatically, and purified water returns to the system or goes on to other uses (a tap is shown on the stand). If the water does not reduce the hardness sufficiently, controlled by a solenoid, it returns to a new cleaning cycle. Thus, a closed cycle enables the water to be purified to a predetermined value (Fig. 1).

[Figure 1 to appear]

**Figure 1.** The water purification machine assembled by schoolchildren.

The idea was so interesting to the students that at the conclusion of the session, the team did not disband. The students created the group NP Team on social media and continue to participate in grant competitions. There is intent to continue the development of a comprehensive method for cleaning contaminated water from CHP plants. The rector of Ural Federal University supported the pilot project, and students were given the opportunity to create prototypes. The only downside is that two of the students are graduating this year. However, one of the graduates is on track to enter Ural Federal University, and the two students of non-graduating classes (40% of the team) are participating in UrFU projects.

The team has also participated in the Ural Project Session in the educational center Sirius (Sochi). One hundred students who had passed competitive selection attended. Students from public, municipal, and private schools located in the Russian Federation in the regions of the Urals and Western Siberia were eligible to participate in the competition. The purpose of the Ural Project Session was the identification, development, and support of gifted students in design and research. Tasks to be performed during the session included:

- activation of creative, cognitive, and intellectual initiatives of students who show interest and an inclination to studying mathematics and earth sciences;

- identification and support of students with an aptitude for research and development;

- generalization and development of best practices in the study of mathematics, physics, chemistry, and biology in upper grades, including preparation for Olympiads, development of research and educational projects, and organization of the extracurricular work of students;

- engagement of scientists, specialists from research institutions, and higher education institutions to work with students.

One of the results was a project to create environmentally friendly thermal power plants using photosynthetic processes. The use of algae to produce future bioethanol fuels is currently very relevant. At the moment, on the basis of the Department of Nuclear Power Plants and Renewable Energy Sources with the help of talented schoolchildren, a project is being developed to obtain bioethanol from algae grown on the polluted waters of thermal power plants. The first experiments conducted by the team showed that the alcohol content in bioethanol from algae grown on water after a contact heat exchanger is 2-2.5 times higher than that of algae grown on pure water. Thus, problems of recycling low-potential waste heat are solved as well as the reduction of carbon dioxide emissions and thermal pollution; ponds accumulators can be converted into bioplants. The use of the photosynthesis process technology will make it possible to significantly compensate for the oxygen consumption of the CHP and TPP.

At the same time, the rate of oxygen generation is 3-5 times higher than the rate in natural ecosystems when using algae and the elevated temperature of the wastewater of thermal power plants. As a result, this will lead to an increase in the quality of life of the population. Thus, a comprehensive reduction of anthropogenic impact, cleaning ponds and obtaining bioethanol will ultimately make CHP and TPP more environmentally friendly.

## 5. Discussion and Conclusions

The growth of the Earth's population, industry development, and an increase in environmental pollution is not a positive prospect for the ecological future of the planet, and the younger generations realize this, the better. Simply contemplating and understanding the scale of the impending disaster is insufficient. To amend the situation, we must create, work on, and try to solve problems. There is an old Soviet film, *Through the Thorns to the Stars*, in which Earthmen visit a planet that was brought to a nearly unfit state by its inhabitants, and who spend their time doing general cleaning. The planet becomes suitable for life again. This is a fantastical dream, but we need to solve the problems ourselves. We argue that the problem can be solved through school and student environmental creativity. Of course, students sometimes do not have sufficient knowledge, but enthusiasm and a desire to apply their knowledge to something important can overcome this. It is probable that this will not influence every student similarly, but the more such programs are available, the faster a generation will develop with a sustainable way of thinking.

It is unjustified to consider that the work of students is insignificant and impractical. There are many ways to reduce emissions at CHPs and TPPs. Nearly all methods of reducing the environmental load are energy-consuming, expensive, and cumbersome, but all were developed in research institutions by large teams. Students do not have the same financial resources, so they use elementary materials in their studies. The outcome is inexpensive and produces a more attractive product, and by solving a specific applied problem, it is, therefore, possible to create an active position about sustainable development in younger generations.

Consider the problem from an institutional theory viewpoint. Normative institutions are among the most rigid formal institutions, which rely on violence as a coercive mechanism. The government relies on this type of institution. Routines are repetitive, normal, and predictable procedures for solving similar problems. The theory of structural institutions was developed at the firm level, but the model also operates at the national level. It determines not only the structure itself but the model of interaction of various elements of a system. Mental institutions or mental models imply a similar perception and interpretation of reality. At the core of mental institutions are values integral to the culture of the individual and society. At the same time, values are divided into both absolutes, which an individual always follows, and relatives, which are conducted only when profitable.

Functional and structural institutions are derived from normative and mental ones, and we can thus define the latter as primary institutions. The importance of mental institutions is greater. Adoption of regulatory institutions is determined by transaction costs of accession and evasion; if transaction costs of evasion are lower than the costs of compliance, a person will choose not to comply with the norm. Another element of public consciousness is ideology, which combined with institutions and mental models determines individual choices and affects economic, social, cultural, and environmental indicators. Mental models, institutions, and ideologies are part of the process by which people interpret and order the environment. Mental models are endogenous and to some extent unique to each person. Exogenous ideologies and institutions provide closer sharing of beliefs and ordering of the environment. Relationships among mental models, ideologies, and institutions depend on the product and the environment. We do not have sufficient information about how progression occurs in mind (Denzau and North, 1994), but we can assert that mental models form most effectively at a young age, and their formation occurs over a long period. In the case of mental institutions, transaction costs of evasion are extremely high because they are supported by both social and internal mechanisms of moral evaluation. The costs of changing mental institutions are also high, and we, therefore, conclude that only people with stable mental models of sustainable development can guarantee a sustainable future. This does not negate the role of normative institutions during the transitional period since the formation of mental institutions takes a long time (at least one or two generations). During this period, normative institutions create greenhouse conditions for the development of mental institutions.

Sustainable Development largely depends on individual competencies. Nevertheless, today there is no agreement among researchers what they should be. This case correlates with the German definition of "shaping competence" (de Haan, 2006; de Haan *et al.*, 2008; Rieckmann, 2012) which includes:

- competency in anticipatory thinking
- competency in interdisciplinary work
- competency in cosmopolitan perception and change of perspectives
- competency in handling incomplete and complex information
- participatory competency
- competency in cooperation
- competency in dealing with individual decision dilemmas
- competency in self-motivation and motivating others
- competency in reflection on individual and cultural models
- competency in independent action
- competency in ethical action
- capacity for empathy and solidarity

How does this skill set correlate with the case described? First and foremost, it is proactive thinking. One should note that the work of large research teams is often somewhat "stamped" and therefore employees are to some extent devoid of creativity. School and university teams in this case, on the contrary, can afford any, sometimes fantastic ideas. If the idea has a further development, then it can develop the competence of advanced thinking. At the same time, unviable ideas evoke a craving for knowledge, a desire to understand why this is impossible now. The project is interdisciplinary. When creating the installation, it was necessary to apply programming skills, design skills, automation skills, and management skills. Each team member was responsible for the separate area, but at the same time, the students worked as one team helping each other. As a result, each of them acquired new competencies and knowledge in other disciplines.

Working on a project that meets one of the global challenges changes perspectives and mindset. The project presents an alternative way to absorb $CO_2$, the main greenhouse gas that poses a threat to all of humanity. Thus, the understanding of the global nature of the task leads to the desire to participate. Everyone participates in the work by virtue of their skills and this creates a team. Having different knowledge and skills, schoolchildren and students completely different in character are to submit their project to the jury during the project session. They can complete the task only in close cooperation and with mutual assistance. We agree that personal background affects performance in ESD (Svanström *et al.*, 2018); at the same time, team work smoothes the differences.

Each member of the team is an equal participant in the research team. This or that task is not imposed on anyone. Everyone chooses a job in a project on which he works independently. The supervisor basically does not offer a ready-made solution, so the team members need to analyze the data themselves and find the missing information. Sometimes, they had to use incomplete data, analyze and interpret it taking into account the capabilities and knowledge to achieve your goal. While working on the interdisciplinary project, everyone is responsible for a particular stage of work: e.g., someone can program, and someone else has design skills. The project approach enables the solution of the problem of interdisciplinarity (Grishaeva *et al.*, 2018; Sinakou *et al.*, 2018) more effectively than the traditional curriculum. Besides, the case supports the thesis that real-world experience is critical for ESD (Molderez and Fonseca, 2018). We had a situation when a member of the team responsible for connecting the automation was able to convince the team to change the design for a more reliable connection of control systems. Thus, the schoolchildren formed the competence of individual decision within the framework of their task.

In the end, the schoolchildren and students working in a team have a different nationality, which leaves an imprint on the behavior of an individual. Teamwork helps to make sense of one's behavior as well to get acquainted with the behavioral models and culture and other team members. In our opinion, the ability to respect others builds the competence of ethical action, i.e. respect is a solution to axiological problems.

However, we must keep in mind that we are considering the initiative of a particular university. Even though such an initiative is not unique, we cannot talk about the existence of an established national policy at the moment. This is what Russia really lacks. In addition, we consider the case from the standpoint of institutional analysis, but Sustainable Development in higher education is an interdisciplinary object of study in its essence. For example, students' perceptions, as well as community engagement, can be the subjects of further case studies and theoretical contributions. Russia, with its transforming higher education system, is an extremely voluminous and significant research niche.

**Acknowledgments:** The authors also wish to express their gratitude to anonymous reviewers, who made this paper much better. The study was conducted with the financial support of the Russian Foundation for Basic Research, project *18-00-01040 KOMFI "The Impact of Emerging Technologies on Urban Environment and the Quality of Life of Urban Communities"*.

**Notes**

[1] An interview was conducted with all participants of the program, and aimed at answering two questions: 1) what is your motive for participating in the program, and 2) what results have you received/plan to get from participation in the program?
[2] A program of the Fund for the Promotion of Innovation that finances youth projects with scientific novelty (http://umnik.fasie.ru).
[3] In 1996, the Vernadsky Fund established a scholarship awarded to environmental students and students of other majors dealing with sustainable development, and in 2004, scholarships for postgraduate and doctoral students were established (http://www.vernadsky.ru/projects-of-the-foundation/scholarships-named).
[4] This major is similar to labor protection.